\documentclass{modelica}
\usepackage{nomencl}
\usepackage{etoolbox}
\renewcommand\nomgroup[1]{%
  \item[\bfseries
  \ifstrequal{#1}{P}{Symbols}{%
  \ifstrequal{#1}{S}{subscript}{%
  \ifstrequal{#1}{O}{Other symbols}{}}}%
]}

\makenomenclature
\hypersetup{%
  pdftitle  = {Latex Template for the International Modelica Conference},
  pdfauthor = {Author Name1, Author Name2},
  pdfsubject = {14th International Modelica Conference 2021},
  pdfkeywords = {Modelica, conference, LaTeX, template},
  colorlinks,
  linkcolor=black,
  urlcolor=black,
  citecolor=black,
  pdfpagelayout = SinglePage,
  pdfcreator = pdflatex,
  pdfproducer = pdflatex}

\usepackage{hyperref}
\begin{document}
\thispagestyle{empty}
\title{Frost/Defrost Models for Air-Source Heat Pumps with Retained Water Refreezing Considered}
\author[1]{Jiacheng Ma}
\author[2]{Matthis Thorade}
\affil[1]{Modelon Inc., USA, {\small\texttt{jiacheng.ma@modelon.com}}}
\affil[2]{Modelon Deutschland GmbH, Germany, {\small\texttt{matthis.thorade@modelon.com}}}


\maketitle\thispagestyle{empty} 
\abstract{%
Cyclic frosting and defrosting operations constitute a common characteristic of air-source heat pumps in cold climates during winter. Simulation models that can capture simultaneous heat and mass transfer phenomena associated with frost/defrost behaviors and their impact on the overall heat pump system performance are of critical importance to improved controls of heat delivery and frost mitigation. This paper presents a novel frost formulation using an enthalpy method to systematically capture all phase-change behaviors including frost formation and melting, retained water refreezing and melting, and water drainage during cyclic frosting and defrosting operations. A Fuzzy modeling approach is proposed to smoothly switch source terms when evaluating the dynamics of frost and water mediums for numerical robustness. The proposed frost/defrost model is incorporated into a flat-tube outdoor heat exchanger model of an automotive heat pump system model to investigate system responses under cyclic operations of frosting and reverse-cycle defrosting. 
}

\noindent\emph{Keywords: heat pump, reverse-cycle defrost, retained water, transient simulation}

\section{Introduction}
Cyclic frost accumulation and removal on the outdoor coil surface of air-source heat pumps are common in winter operations. As the system performance degrades due to blockage of free air flow passages and increased thermal resistance between coil walls and air streams, periodic defrost cycles are necessary to ensure consistent heat delivery and safe operations of the system. In practice, reverse-cycle defrost is among the most common strategies to implement by reversing the thermodynamic cycle to melt the accumulated frost. After that, the melted water can be evaporated as the coil continues being heated and can simultaneously drain from the surface due to gravity. Once the defrost cycle completes according to feed-back controls or simply time-based controls, the refrigerant flow is reversed again to restore nominal heating operations. However, in case of an insufficient defrost operation, a portion of the melted liquid water is retained on the coil surface owing to surface tensions, which subsequently refreezes as solid ice when the system switches back to the heating mode and the outdoor coil operates as an evaporator. As a consequence, the next heating cycle can suffer from more significant performance degradation due to the presence of the ice layer as additional thermal resistance compared to the previous heating cycle, which can lead to a shorter heating period before a necessary defrost cycle \cite{zhang2024performance}. The issue is particularly important to operations of microchannel heat exchangers commonly adopted in automobile applications (including electric vehicles) whose performance is more sensitive to air flow blockage and thermal insulation due to the compact configuration. 

Simulation tools that can capture the complicated heat and mass transfer phenomena of cycling frosting-defrosting operations can be critically useful for the design and evaluation of improved defrost controls for air-source heat pumps that can mitigate frost/ice on the outdoor coil without excessively heating the coil during defrost cycles. Almost all existing frost/defrost models for heat pump systems focused on switches between frost formation and melting behaviors, omitting the impacts of retained water refreezing and melting on transients of cyclic frosting and defrosting \cite{qiao2018modeling,ma2023development,alam2024}. \textcite{westhaeuser2023model} derived a numerical model that accounted for frost formation and melting as well as retained water statuses in a lumped control volume. However, no general switching scheme was implemented for transitions between stages of frost and water processes, which hinders its applicability to describe heat pump transients under realistic mode switches between frosting and defrosting. 

This paper presents a novel frost model that systematically captures all phase-change behaviors of frost and retained water/ice in a multi-phase multi-substance control volume using an enthlapy method. A Fuzzy modeling approach is proposed to smoothly switch source terms of governing equations to evaluate dynamics of the control volume based on underlying physical processes. The derived model is incorporated into a discretized flat-tube outdoor heat exchanger model of an automotive heat pump system. The system model is then simulated to investigate transients under cyclic operations of frosting/heating and reverse-cycle defrosting with retained water behaviors taken into consideration. 

The remainder of this paper is organized as follows: Section \ref{sec: frost model} describes model development for frost and retained water processes; Section \ref{sec: heat pump model} presents a discretized outdoor heat exchanger model incorporated with frost behaviors, and other component models to complete a heat pump system model; simulation results of the developed system model under cyclic frosting and defrosting are reported in Section \ref{sec: simulation results}, followed by conclusions summarized in Section \ref{sec: conclusion}.

\begin{figure*}[thb]
\centering
\includegraphics[width=0.7 \textwidth]{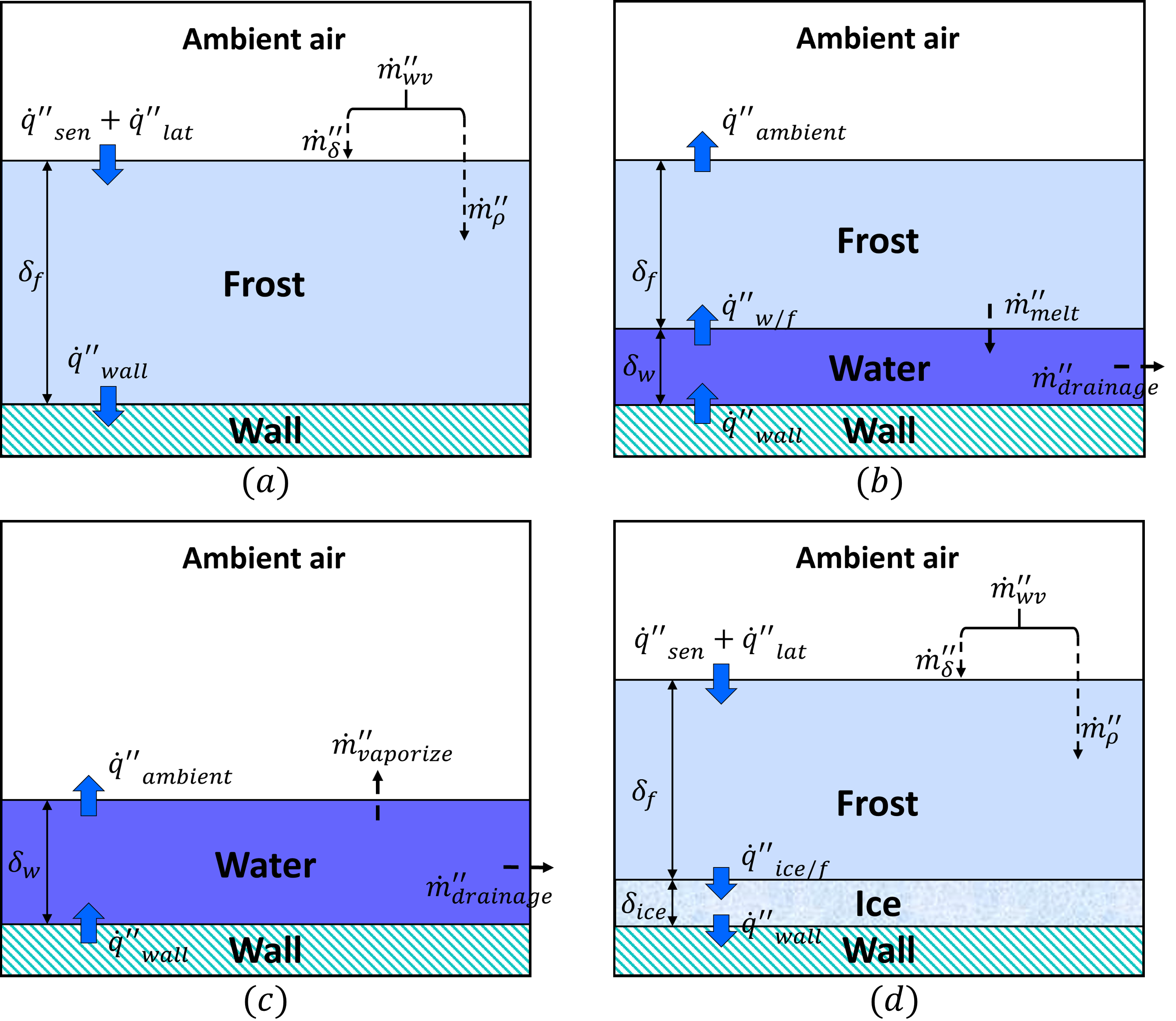}
\caption{Frost and water/ice processes during cyclic frosting and defrosting: (a) frost formation on dry coils; (b) frost melting and water drainage; (c) retained water vaporization and drainage; (d) retained water refreezing and frost formation on ice.}
\label{fig: frost processes}
\end{figure*}

\section{Frost/defrost model}\label{sec: frost model}
Frost formation and melting involve simultaneous heat and mass transfer. A common practice to model both processes relies on evaluating mass and energy balances of a frost control volume to characterize its dynamics which can be represented by a frost layer thickness and a lumped density. While a control volume for water/ice can be set up in a similar manner and interconnected with the frost volume to capture retained water/ice behaviors, significant model complexities are introduced as different phase presences and interactions between these control volumes need to be captured when the operating mode changes. Figure \ref{fig: frost processes} depicts typical processes that occur on coil wall surfaces under cyclic frosting and defrosting. Clearly, each stage during the process corresponds to different heat and mass transfer descriptions for control volumes and interfaces. Moreover, additional iteration variables (e.g., interface temperature or heat flux) can be introduced when modeling frost and water/ice mediums with individual control volumes. As a consequence, the overall model robustness deteriorates. 

This work aims to develop a frost model with retained water/ice considered in a uniform model structure. A lumped film control volume that integrates frost, liquid water, and solid ice is developed using an enthalpy method \cite{alebrahim2002electric}. The film dynamics under various operating modes and phase presences are captured by switching source terms of mass and energy balances.

\subsection{A lumped film volume model}
An advantage of the enthalpy method to model multiple mediums with phase change is that phase presences can be determined using the total enthalpy and mass fractions of each medium, eliminating the evaluation of internal heat transfer between them explicitly. Consider a one-dimensional lumped film control volume composed of frost and retained water/ice
\begin{align}
    m_{film} = m_f+m_{wi}.
\end{align}

The total enthalpy of the film is defined as
\begin{align}\label{eq: film enthalpy}
 H_{film}=m_fh_f+m_{wi}h_{wi}
\end{align}
which is the sum of frost enthalpy and retained water/ice enthalpy. Carrying out an energy conservation analysis of the film volume yields
\begin{align}\label{eq: energy conservation}
    \frac{dH_{film}}{dt}=\dot{H}_{in} - \dot{H}_{out} + \dot{Q}
\end{align}
where the inlet enthalpy flow $\dot{H}_{in}$ corresponds to water vapor desublimation during frost formation, the exit enthalpy flow $\dot{H}_{out}$ corresponds to melted water drainage and vaporization during frost melting, $\dot{Q}$ represents heat transfer rate (convection or conduction) with ambient air and coil metal walls. It is important to note that the enthalpy flow source terms are only active corresponding to a specific operating mode (e.g, frost formation, frost melting), and the heat transfer rate changes signs when the heat flow directions reverse under mode switches.

Assume that the mediums of the film volume are in thermal equilibrium and their temperature gradients are negligible. Therefore, a lumped temperature can be determined from the enthalpy. Note that frost is a mixture of dry air and ice. Equation \ref{eq: film enthalpy} can be rewritten as
\begin{align}
    H_{film}=m_{ice}h_{ice}+m_{air}h_{air} + m_{wi}h_{wi}
\end{align}
where $m_{ice}$ and $m_{air}$ denote mass of ice and air of the frost layer, respectively. Apparently, the determination of the film temperature depends on operating modes. In case of frost formation, the retained water on the coil surface must be frozen before frost deposition happens, since a necessary condition of frost formation is that the surface temperature is below the water freezing point $T_0$ (e.g., 273.15 K). This physical insight indicates that the film undergoes frost formation process when the condition $H_{film}<H_{film,s0}$ holds where $H_{film,s0}$ represents the enthalpy when all retained water turns into solid ice at $T_0$ and is computed by
\begin{align}
    H_{film,s0}=m_{ice}h_{ice,0}+m_{air}h_{air,0}+m_{wi}h_{ice,0}
\end{align}
where $h_{ice,0}$ and $h_{air,0}$ represent specific enthalpies at the freezing temperature. During frost formation, the film enthalpy is proportional to temperature changes with a lumped specific heat capacity since there is no internal phase change inside the volume. As a result, the film temperature is obtained by
\begin{align}\label{eq: solid temperature}
    T_{film}=\frac{H_{film}-H_{film,s0}}{m_{film}c_{p,film}}+T_0
\end{align}
where $c_{p,film}$ is a mass-weighted property and is determined by
\begin{align}
    c_{p,film}=\frac{c_{p,ice}m_{ice}+c_{p,air}m_{air}+c_{p,wi}m_{wi}}{m_{film}}.
\end{align}

Similarly, an enthalpy condition $H_{film}>H_{film,l0}$ can be employed to indicate the conclusion of the frost melting process as well as the switch to retained water drainage and vaporization as shown in Figure \ref{fig: frost processes} (c) where
\begin{align}
    H_{film,l0}=m_{ice}h_{w,0}+m_{air}h_{air,0}+m_{wi}h_{w,0}
\end{align}
defines a film enthalpy at $T_0$ when solely liquid water is present in the film volume. In this case, further heating of the film leads to a temperature above the freezing point, which can be determined using Equation \ref{eq: solid temperature} with $H_{film,l0}$ as a reference. 

Clearly, the film undergoes phase changes associated with either the frost layer or the retained water/ice when $H_{film,s0}<H_{film}<H_{film,l0}$ is met, with a temperature at $T_0$. These processes include frost melting, and retained water refreezing or melting.

In summary, the film temperature can be retrived from its enthalpy and mass composition at any time instance as
\begin{align}
    T_{film} = \begin{cases}
        \frac{H_{film}-H_{film,s0}}{m_{film}c_{p,film}}+T_0 & \text{if}\;H_{film}<H_{film,s0},\\
        \frac{H_{film}-H_{film,l0}}{m_{film}c_{p,film}}+T_0 & \text{if}\;H_{film}>H_{film,l0},\\
        T_0 & \text{otherwise}.
    \end{cases}
\end{align}

In terms of mass conservation, analyses are carried out for the frost layer and the retained water/ice layer separately. The frost mass is characterized by its thickness and density $m_f=\delta_f\rho_fA_s$ as mass transfer contributes to its growth and densification simultaneously. The mass conservation can then be formed as 
\begin{align}
    \frac{d\delta_f}{dt}& =\frac{\dot{m}_{\delta,in}-\dot{m}_{\delta,out}}{\rho_fA_s},\label{eq: frost thickness}\\
    \frac{d\rho_f}{dt}& =\frac{\dot{m}_{\rho,in}-\dot{m}_{\rho,out}}{\delta_fA_s}\label{eq: frost density}
\end{align}
with a decomposition of the source mass flow rates into portions that correspond to mass transfer at the frost surface and within the frost layer. Finally, a mass balance of the retained water/ice layer is formed as
\begin{equation}
    \frac{dm_{wi}}{dt}=\dot{m}_{wi,in}-\dot{m}_{wi,out}.
\end{equation}

With the model setup presented above, transient behaviors of the lumped film can be described by its enthalpy, the frost layer thickness and density, and the retained water/ice mass. Other properties of the film are retrieved from these state variables. For example, an average film density is computed by
\begin{align}
    \rho_{film}=\frac{m_{film}}{\delta_f+m_{wi}/\rho_{wi}}
\end{align}
where $\rho_{wi}$ represents water/ice density depending on the operating mode.

\subsection{Switched source terms}
\subsubsection{Heat and mass transfer}\label{sec: heat and mass source terms}
Evaluations of the source terms that appear in the governing equations depend on operating modes and phase presences of the film. In the current work, coupled behaviors of the frost layer and water/ice layer are described in the following stages which cover scenarios associated with cyclic frosting and defrosting operations of heat pumps: 
\begin{enumerate}
    \setlength{\itemsep}{-1pt}
    \item frost formation on retained solid ice, as shown in Figure \ref{fig: frost processes} (d);
    \item melting of the retained ice and frost, as shown in Figure \ref{fig: frost processes} (b);
    \item retained liquid water drainage and vaporization, as shown in Figure \ref{fig: frost processes} (c);
    \item dry coil with no frost or retained water/ice.
\end{enumerate}

To derive source terms of heat and mass transfer under frost formation at stage 1, an important assumption is that water vapor is saturated at the frost surface. The mass flow rate of water vapor that diffuses into the frost layer and contributes to its densification over the surface area $A_s$ can be obtained according to Fick's law at the frost surface
\begin{align}
    \dot{m}_{\rho,in}=D_{eff}A_s\left. \frac{\partial \rho_{wv}}{\partial x}\right|_{x=\delta_f}
\end{align}
and is approximated in a finite-difference fashion across the film volume
\begin{align}
    \dot{m}_{\rho,in}\approx D_{eff}A_s\frac{\rho_{wv}(T_{fs})-\rho_{wv}(T_{wall})}{\delta_f}
\end{align}
where $\rho_{wv}(T_{fs})$ and $\rho_{wv}(T_{wall})$ denote saturated water vapor density at the film surface and wall temperatures, respectively. The effective diffusion coefficient $D_{eff}$ is evaluated using the following correlation \cite{breque2016frosting}
\begin{align}
    D_{eff}=\frac{\epsilon}{1-0.58(1-\epsilon)}
\end{align}
where $\epsilon$ represents the frost porosity. Given a total water vapor mass flow rate from air streams $\dot{m}_{wv}$, the portion that deposits on the frost surface can be obtained by
\begin{align}
    \dot{m}_{\delta,in}=\dot{m}_{wv}-\dot{m}_{\rho,in}. 
\end{align}

Since there is no mass flow leaving the frost layer during this stage, Equation \ref{eq: frost thickness} and \ref{eq: frost density} can be evaluated with $\dot{m}_{\delta,in}$ and $\dot{m}_{\rho,in}$, respectively. Furthermore, the retained water/ice layer yields no mass transfer, which leads to the mass $m_{wi}$ remaining constant. 

Heat transfer between the film volume and air streams consists of sensible and latent heat due to water vapor desublimation. Carrying out an energy balance at the film surface leads to 
\begin{gather}
    \dot{Q}_a=A_sk_{film}\frac{T_{fs}-T_{film}}{\delta_{film}/2}
\end{gather}
which can be used to obtain the film surface temperature $T_{fs}$. $k_{film}$ is the film thermal conductivity which is determined in a density-weighted manner
\begin{gather}
    k_{fim}=k_f+(k_{wi}-k_f)\frac{\rho_{film}-\rho_f}{\rho_{wi}-\rho_f}.
\end{gather}
A correlation reported in \cite{lee1997one} is adopted to evaluate the frost thermal conductivity as a function of the frost density. Similarly, heat transfer from wall surfaces is calculated by
\begin{align}\label{eq: film wall heat transfer}
    \dot{Q}_{wall}=A_sk_{film}\frac{T_{wall}-T_{film}}{\delta_{film}/2}.
\end{align}

As a result, the energy conservation in Equation \ref{eq: energy conservation} becomes 
\begin{align}\label{eq: energy conservation stage 1}
    \frac{dH_{film}}{dt}=\dot{Q}_a+\dot{Q}_{wall}
\end{align}
for this stage. Note that the inlet enthalpy flow carried by water vapor is negligible compared to sensible and latent heat which is determined in an air flow model presented in the next section.

At stage 2, the retained ice is melted, followed by the frost melting process. To calculate the melting rate, a quasi-steady-state energy balance is employed
\begin{gather}
    \dot{m}_{melt}=\frac{\dot{Q}_a+\dot{Q}_{wall}}{\Delta h_{sl}}
\end{gather}
where $\Delta h_{sl}$ represents the water heat of fusion. Typically, heat exchanger fans are off under defrost mode. Therefore, the only mass transfer source of the frost layer is the melted water $\dot{m}_{\delta,out}=\dot{m}_{melt}$, which flows into the retained water/ice layer and implies $\dot{m}_{wi,in}=\dot{m}_{melt}$. Furthermore, it is assumed that the frost density remains constant during frost melting. The energy conservation shown in Equation \ref{eq: energy conservation stage 1} also applies to this stage. 

After frost is melted and with liquid water retained on coil surfaces, stage 3 focuses on transients of the retained water. Two primary processes are liquid drainage owing to gravity and vaporization as heating coil surfaces leads to mass transfer. It is challenging to quantify the water drainage rate physically since it relies on a number of factors such as surface orientation, wettability, etc. In the current work, a first-order model 
\begin{align}
    \dot{m}_{drain}=\frac{m_{wi}}{t_{drain}}
\end{align}
is applied to describe the process \cite{westhaeuser2023model}. The drainage flow rate is proportional to the total retained water mass with a time constant $t_{drain}$ which needs to be calibrated for a specific application. 

The mass flow rate of vaporized water is determined by
\begin{align}
    \dot{m}_{vap}=\alpha_m\big(\rho_{wv}(T_{film})-\rho_{wv}(T_a)\big)
\end{align}
where $\alpha_m$ is a mass transfer coefficient, $\rho_{wv}(T_{film})$ is the saturated water vapor density evaluated at the film temperature, and $\rho_{wv}(T_a)$ is the water vapor density at the ambient temperature. Since both drainage and vaporization move water out of the film control volume, the final form of mass balance for this stage is
\begin{align}
    \frac{dm_{wi}}{dt}=-\dot{m}_{drain}-\dot{m}_{vap}.
\end{align}

Correspondingly, enthalpy flows that result from the retained water mass transfer are evaluated in energy conservation. The outlet enthalpy flow is determined by
\begin{align}
    \dot{H}_{out}=\dot{m}_{drain}h_{film}+\dot{m}_{vap}\Delta h_{lv}
\end{align}
where $h_{film}$ denotes the film specific enthalpy and $\Delta h_{lv}$ denotes the water heat of vaporization.

In case the retained water is completely removed from coil surfaces before a switch back to frosting conditions, heat is dissipated to the ambient air. At stage 4, the film energy balance is describe by Equation \ref{eq: energy conservation stage 1} without mass transfer. 

\subsubsection{Switching algorithm}
Source terms associated with all stages are computed at each time step. However, only those relevant to the current stage should be applied in evaluating the governing equations. A robust switching algorithm is crucial for realizing the proposed film model. In the current work, the T-S Fuzzy modeling approach \cite{takagi1985fuzzy} is adopted for this purpose. 

\begin{figure}[ht]
\centering
\includegraphics[width=0.45 \textwidth]{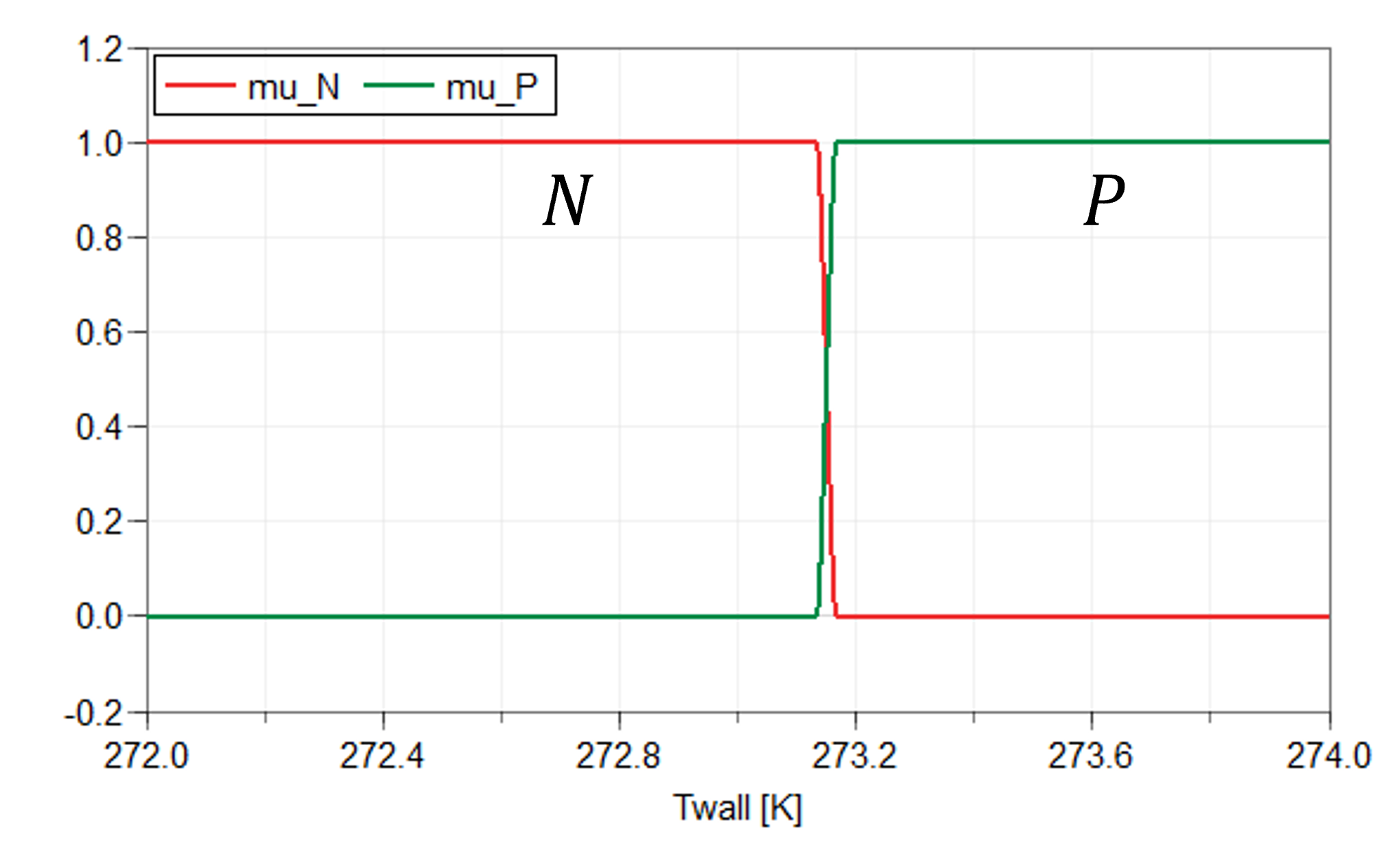}
\caption{Membership functions associated with Fuzzy numbers for the linguistic variable $T_{wall}$.}
\label{fig: fuzzy number}
\end{figure}

A set of switching rules is first introduced based on characteristics of all possible stages described in Section \ref{sec: heat and mass source terms}. Apparently, the wall temperature constitutes a candidate for differentiating the frost formation process from other melting and dry heating stages. Specifically, stage 1 is active when the wall temperature is below the freezing point temperature $T_0$. Otherwise, one of the stages during defrosting is active. With frost presence, the film volume undergoes stage 2 until all the frost is melted. Numerically this can be determined when the frost layer thickness is above a small threshold (for model robustness concerns). Similarly, the retained water presence is indicated by the retained water/ice layer mass, that also determines whether stage 3 or stage 4 is active. In summary, wall temperature $T_{wall}$, frost layer thickness $\delta_f$, and retained water/ice layer mass $m_{wi}$ are identified as linguistic variables to construct Fuzzy rules. For each variable, two Fuzzy numbers are created with unique membership functions. For instance, Figure \ref{fig: fuzzy number} depicts membership functions associated with Fuzzy numbers $N$ and $P$ for the linguistic variable $T_{wall}$. It can be seen that values of both membership functions change around the freezing point temperature $T_0$ which corresponds to rules that depend on whether $T_{wall}$ is above or below $T_0$. It is important to note that the piecewise membership functions are continuous and differentiable to avoid generating events during simulations. Fuzzy numbers associated with $\delta_f$ and $m_{wi}$ are acquired similar to those shown in Figure \ref{fig: fuzzy number} where membership functions associated with Fuzzy numbers $N$ and $P$ change values between 0 and 1 around a threshold. In this case, $\delta_{f,\epsilon}=0.01 \mathrm{mm}$ and $m_{wi,\epsilon}=1e^{-5} \mathrm{kg}$ are applied as thresholds.

After that, the following set of Fuzzy rules can be formed using Fuzzy numbers
\begin{enumerate}
\setlength{\itemsep}{-1pt}
    \item stage 1 IF $T_{wall}$ is $N$;
    \item stage 2 IF $T_{wall}$ is $P$ AND $\delta_f$ is $P$;
    \item stage 3 IF $T_{wall}$ is $P$ AND $\delta_f$ is $N$ AND $m_{wi}$ is $P$;
    \item stage 4 IF $T_{wall}$ is $P$ AND $\delta_f$ is $N$ AND $m_{wi}$ is $N$,
\end{enumerate}
to identify continuous mode switches under cyclic operations. Finally, numerical weights are assigned to source terms associated with each stage to realize mode switches between these stages through a defuzzification process revealed below
\begin{align}
    \omega_1& =\mu_N(T_{wall}),\\
    \omega_2& =\mu_P(T_{wall})\times \mu_P(\delta_f),\\
    \omega_3& =\mu_P(T_{wall})\times \mu_N(\delta_f)\times \mu_P(m_{wi}),\\
    \omega_4& =\mu_P(T_{wall})\times \mu_N(\delta_f)\times \mu_N(m_{wi}).
\end{align}
which translates the Fuzzy rules to numerical outputs using membership function values. A weight for each stage is then obtained through normalization
\begin{align}
    \gamma_i=\frac{\omega_i}{\sum_{i=1}^4\omega_i}.
\end{align}

The final set of governing equations for the film model is presented as
\begin{align}
    \frac{dH_{film}}{dt}&=\dot{Q}_a + \dot{Q}_{wall} - \gamma_3(\dot{m}_{drain}h_{film}+\dot{m}_{vap}\Delta h_{lv})\\
    \frac{d\delta_f}{dt}&=\frac{\gamma_1\dot{m}_{\delta,in}-\gamma_2\dot{m}_{melt}}{\rho_f A_s}\\
    \frac{d\rho_f}{dt}&=\frac{\gamma_1\dot{m}_{\rho,in}+(\gamma_3+\gamma_4)(\rho_{reset}-\rho_f)/t_{reset}}{\delta_f A_s}\label{eq: frost density dynamics}\\
    \frac{dm_{wi}}{dt}&=\gamma_2\dot{m}_{melt}-\gamma_3(\dot{m}_{drain}+\dot{m}_{vap}).
\end{align}

Note that in Equation \ref{eq: frost density dynamics} the second source term is a numerical treatment using a first-order filter to reset the frost density to an initial value for the next frosting cycle after all the frost is melted. It is necessary for simulations of cyclic frost behaviors since the assumption of a constant frost density is applied to model the frost melting process (stage 2). Alternatively, this can be realized using the reinit() method of the Modelica language. 

\section{Reversible heat pump system model}\label{sec: heat pump model}
The developed film volume model captures transient behaviors of frost and retained water/ice on coil surfaces. Incorporating the film model into an outdoor coil model for a reversible heat pump  enables investigation of system performance under cyclic frosting and defrosting operations. This section describes a discretized outdoor heat exchanger model, and then presents other component models as well as system model integration. 
\subsection{Flat-tube heat exchanger model}
\begin{figure}[h]
\centering
\includegraphics[width=0.4 \textwidth]{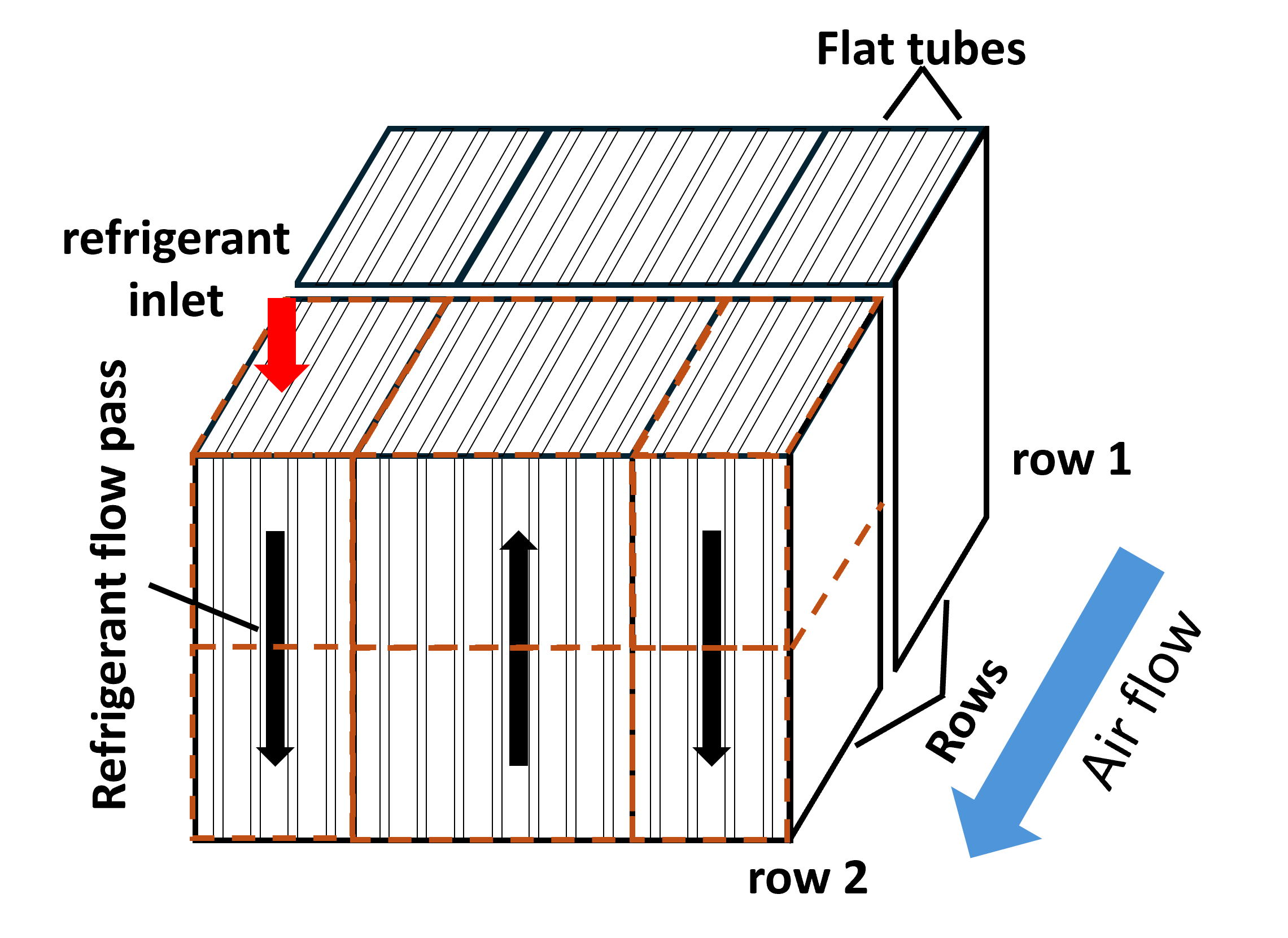}
\caption{Flat-tube heat exchanger discretization.}
\label{fig: heat exchanger discretization}
\end{figure}

An outdoor flat-tube heat exchanger with 6 refrigerant flow passes arranged in 2 rows is discretized by dividing each refrigerant flow pass into two volumes and dividing the air flow path by tube rows. As a result, the heat exchanger is divided into a total number of 12 control volumes. Figure \ref{fig: heat exchanger discretization} depicts the heat exchanger configuration and discretization for a tube row. 

Within each control volume, the reversible refrigerant dynamics are described by the following mass and energy balances
\begin{align}
    \frac{d(\rho_r V)}{dt}&=\dot{m}_{in}-\dot{m}_{out}\\
    \frac{d(\rho_r Vu_r)}{dt}&=\dot{m}_{in}h_{in}-\dot{m}_{out}h_{out}-\dot{Q}_r
\end{align}
along with a steady-state momentum balance that correlates the refrigerant pressure drop with wall frictions. The heat transfer rate to metal walls is obtained by
\begin{align}
    \dot{Q}_r=\alpha_rA_{wall}(T_r-T_{wall})
\end{align}
where $\alpha_r$ is the heat transfer coefficient evaluated by an empirical correlation. The refrigerant pressure and enthalpy are selected as state variables, and all other thermophysical properties (e.g., $T_r$, $u_r$) are retrieved based on them with lookup table interpolation property models. It is important to note that the refrigerant flow descriptions are valid for arbitrary flow directions and handle reverse flow in a numerically sound way. Since the focus of this paper is on frost/defrost model formulations, readers are referred to \cite{ma2024development} for two-phase flow modeling details. 

Metal walls of flat tubes and fins are modeled as lumped capacitance 
\begin{align}
    \frac{dT_{wall}}{dt}=\dot{Q}_r-\dot{Q}_{wall}
\end{align}
where $\dot{Q}_{wall}$ is the heat transfer to the film and is defined in Equation \ref{eq: film wall heat transfer}.

The air-side heat and mass transfer over the film are modeled with a quasi-steady-state assumption due to their fast responses compared to the refrigerant flow or frost. A mass balance is formed as
\begin{align}
    \dot{m}_{in}-\dot{m}_{out}-\dot{m}_{cond}=0
\end{align}
where $\dot{m}_{cond}$ represents the water vapor mass transfer through condensation or desublimation, and is obtained by
\begin{gather}
    \dot{m}_{cond}=\alpha_m A_s(\omega_{in}-\omega_{sat}(T_{fs}))
\end{gather}
where $\alpha_m$ is a mass transfer coefficient evaluated using the heat and mass transfer analogy with Lewis number $Le=0.89$, $\omega_{in}$ represents the inlet air humidity ratio, and $\omega_{sat}(T_{fs})$ represents the saturated air humidity ratio evaluated at the film surface temperature. The energy balance is formed as
\begin{align}
    \dot{H}_{in} - \dot{H}_{out} - \dot{Q}_a = 0
\end{align}
where $\dot{Q}_a$ represents the total heat transfer rate that consists of sensible and latent parts, and is calculated by
\begin{align}
    \dot{Q}_a=\alpha_aA_s(T_a-T_{fs})+\dot{m}_{cond}\Delta h_{lat}.
\end{align}
Here $\alpha_a$ represents the air-side heat transfer coefficient and $\Delta h_{lat}$ is the latent of either vaporization or sublimation depending on the surface temperature. Note that the outlet enthalpy flow $\dot{H}_{out}$ includes those of the air stream as well as the condensated water. A lumped air-side pressure loss is calculated based on the outlet air of the heat exchanger instead of air streams in each channel. 

\begin{figure*}[thb]
\centering
\includegraphics[width=0.8 \textwidth]{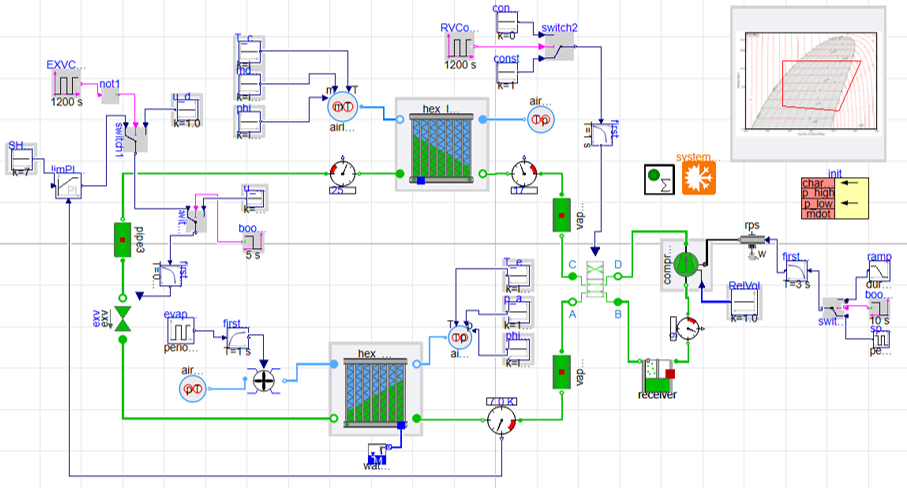}
\caption{Heat pump system model schematic.}
\label{fig: system model}
\end{figure*}

\subsection{Heat pump system model integration}
The developed flat-tube outdoor heat exchanger model incorporated with frost/defrost is subsequently integrated into an automotive air-source heat pump system model. The heat pump can operate in heating or cooling/defrost modes by controlling a reversing valve. R1234yf is utilized as the working fluid. The system adopts a variable-speed compressor and an electronic expansion valve (EXV) as the expansion device. Moreover, an indoor flat-tube heat exchanger that is similar to the outdoor one but with different geometries is employed. 

Modeling the indoor heat exchanger follows the same approach for the refrigerant flow, metal walls, and air flows as presented before. The only distinction compared to the outdoor heat exchanger model is the absence of the film model since no frost/defrost phenomenon occurs on the indoor side. 

A semi-empirical compressor model is employed to calculate the compressor mass flow rate as well as discharge enthalpy using volumetric and isentropic efficiencies which are correlated as functions of discharge and suction pressures and the compressor speed. Furthermore, the shaft power is determined using the shaft torque and speed. 

The EXV model follows a standard throttling process for both flow directions. The mass flow rate across the EXV is computed by
\begin{gather}
    \dot{m}=K_v\sqrt{\rho_{in}|p_{in}-p_{out}|}
\end{gather}
where $K_v$ is a flow coefficient determined by lookup tables when the valve position changes to capture the valve characteristics under varying operating conditions. 

The outdoor heat exchanger fan is modeled using performance tables to calculate the nominal fan power as well as volumetric flow rate that the fan delivers based on the pressure rise across the fan, which is equalized to the pressure drop of the heat exchanger. Then the actual performance of the fan is obtained by scaling nominal values with the fan speed. 

The suction accumulator captures the refrigerant mass and energy conservation, similar to the heat exchanger model. However, the refrigerant exit quality is determined as a function of the liquid level. 

Actuation signals for the heat pump system model include the compressor speed, outdoor fan speed, EXV opening, indoor air flow rate, and the reversing valve opening which controls the refrigerant flow direction and the system operating mode. A schematic of the heat pump system model is shown in Figure \ref{fig: system model}. Besides the major components described above, the system model also include connecting pipes, sensors, and controls. All component models are implemented using Modelon's Air Conditioning Library (\url{https://modelon.com/library/air-conditioning-library/}). 

\section{Simulation results}\label{sec: simulation results}
The developed heat pump system model is simulated to investigate system performance under cyclic frosting and reverse-cycle defrosting operations and to demonstrate capabilities of the developed film model in capturing frost and retained water/ice behaviors. Simulations are carried out using Dymola with the Dassl solver and a tolerance of $1e^{-4}$.

The system is initialized at an off condition where the refrigerant pressure is equalized across the system. Vapor refrigerant resides in the indoor heat exchanger while two-phase refrigerant is stored in the outdoor heat exchanger which is in thermal equilibrium with the ambient. Upon a heating cycle startup, the compressor speed, outdoor fan speed, and indoor air mass flow rate are ramped up to their nominal values while the EXV opening is fixed. After the compressor speed settles, superheat control is activated to regulate the EXV opening using a PI controller, as shown in Figure \ref{fig: system model}. After a period of operations under frosting conditions, the system switches to a defrost cycle to remove the accumulated frost and restore system performance. A number of actuations are considered during mode switching. The EXV is fully opened while the compressor operates at a lower speed than the nominal value during defrost. In the meantime, the reversing valve is energized to redirect the refrigerant flow. On the air side, the outdoor fan is shut off. However, the indoor air mass flow rate remains the same as before in order to heat the refrigerant that flows back to the compressor in the reversed direction. Similarly, when the system switches back to the heating mode, the reversing valve is energized again and other actuators follow the same operations as heating cycles. 

In the current work, the indoor air condition is set to 20 $\mathrm{C}^{\circ}$ with a relative humidity of 50\%, while the outdoor ambient condition is 2 $\mathrm{C}^{\circ}$ with a relative humidity of 70\%. A simulation concerning 25-minute operations is carried out where the system operates in heating mode for 20 minutes before switching to a short defrost cycle, which lasts for 1 minute. After that, the system switches back to the heating model and runs for 4 minutes. 

\begin{figure}[h]
\centering
\includegraphics[width=0.45 \textwidth]{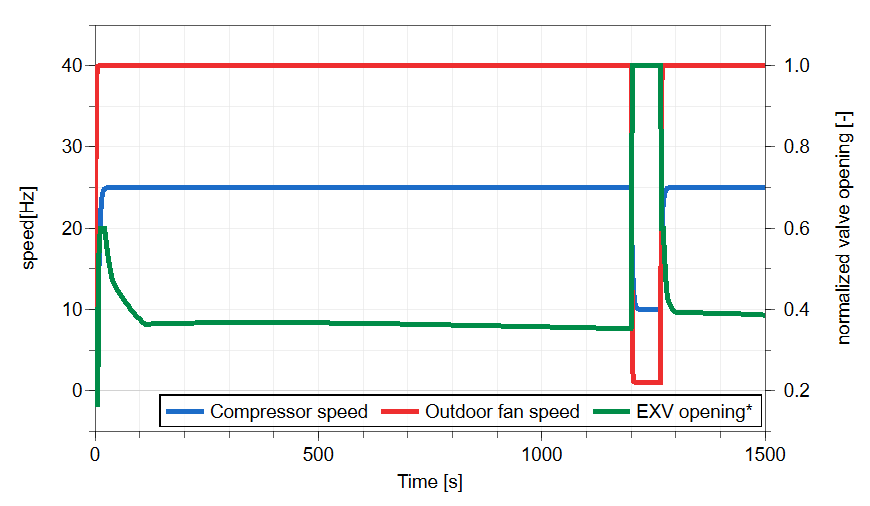}
\caption{Actuation signals of the compressor and outdoor fan speeds and normalized EXV opening.}
\label{fig: actuation}
\end{figure}

\begin{figure}[h]
\centering
\includegraphics[width=0.45 \textwidth]{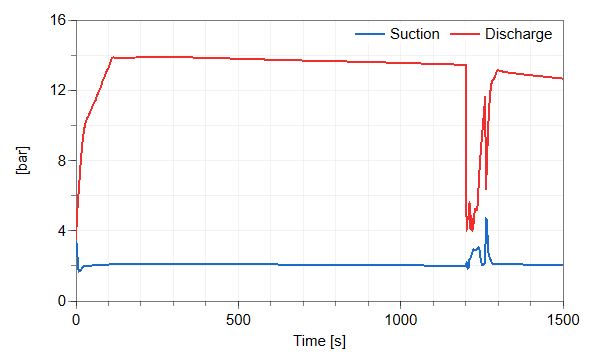}
\caption{Simulation results of discharge and suction pressures.}
\label{fig: pressure}
\end{figure}

Figure \ref{fig: actuation} reveals the compressor speed, outdoor heat exchanger fan speed, and EXV opening corresponding to the operating characteristics described above. Figure \ref{fig: pressure} reports discharge and suction pressures. It can be seen that the discharge pressure settles at around 14 bar while the suction pressure at 2 bar after a startup period of 100 seconds. Then a gradual decrease of the discharge pressure can be observed during the heating cycle. This is due to frost formation on the outdoor heat exchanger that deteriorates heat transfer and leads to a slight decrease of the evaporating temperature. A dramatic drop of the discharge pressure at 1200 seconds is attributed to the actuation of the reversing valve and opening of the EXV. Ater the refrigerant flow is reversed, pressure difference across the compressor starts building up again similar to a startup operation until the reversing valve is energized to conclude the defrost cycle, which is followed by another heating cycle with notable performance degradation due to frost accumulation. 

\begin{figure}[h]
\centering
\includegraphics[width=0.45 \textwidth]{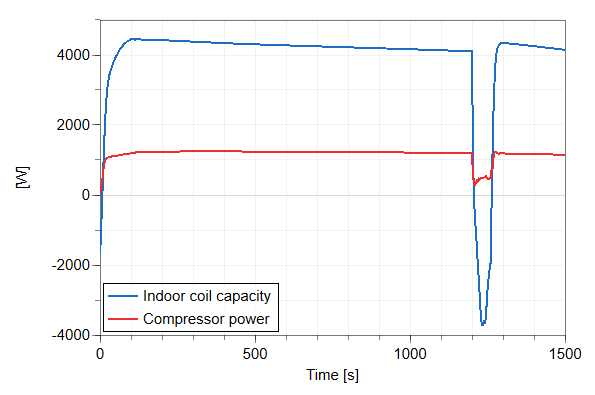}
\caption{Simulation results of indoor air-side capacity and compressor power.}
\label{fig: capacity}
\end{figure}

The same trend can be observed in the indoor heating capacity and the compressor power, as shown in Figure \ref{fig: capacity}. Clearly, the frost buildup leads to a reduction of the air-side capacity during the heating cycle. However, as a result the compressor power also decreases due to a smaller pressure ratio. During the defrost cycle, the indoor capacity changes sign since the indoor heat exchanger operates as an evaporator, which indicates that the heat pump system absorbs heat from the indoor environment and can cause thermal discomfort. 

\begin{figure}[h]
\centering
\includegraphics[width=0.45 \textwidth]{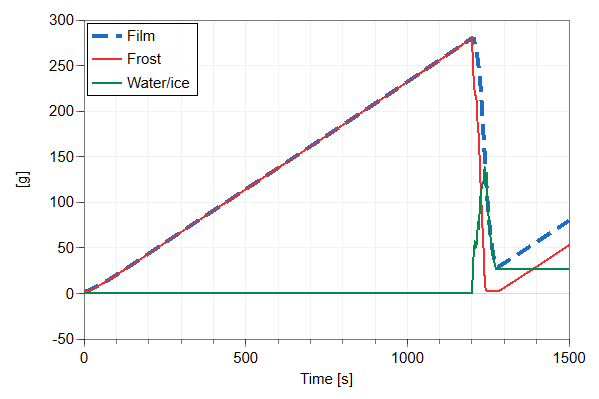}
\caption{Simulation results of frost and retained water/ice mass.}
\label{fig: film mass}
\end{figure}

Figure \ref{fig: film mass} examines evolution of the frost mass and the retained water/ice mass throughout cyclic operations, as a major contribution of the present work is to derive frost models with retained water/ice behaviors taken into consideration. The film mass as the sum of them is also shown. The frost mass keeps increasing upon system startup, while no mass transfer associated with the retained water/ice occurs during this period. The accumulated frost is melted during the defrost cycle, which leads to an increased retained water mass. Meanwhile, the retained water drains away and vaporizes as coil wall surfaces are heated. However, the system switches back to the heating mode before a complete removal of the retained water. Consequently, the retained water refreezes on coil surfaces as frost starts buildup in the next heating cycle. This additional thermal resistance added on the heat exchanger surface and blockage of free air flow passages can accelerate system performance deterioration in the next heating cycle. These results demonstrate that the proposed model is capable of capturing the retained water refreezing phenomenon under cyclic operations. 

\begin{figure}[h]
\centering
\includegraphics[width=0.45 \textwidth]{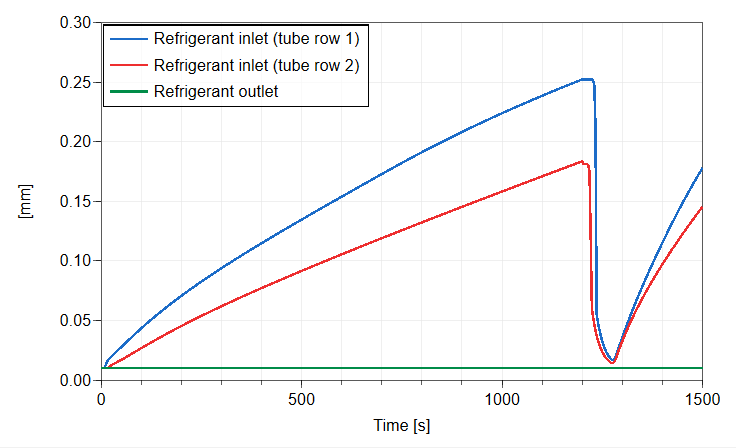}
\caption{Simulation results of discretized film thicknesses of selected control volumes.}
\label{fig: thickness}
\end{figure}

Furthermore, Figure \ref{fig: thickness} reveals non-uniform film thicknesses across the outdoor heat exchanger. Film thicknesses of control volumes at the refrigerant flow inlet (the first control volume of tube row 2), in the middle of the refrigerant path (the first control volume of tube row 1), and at the refrigerant flow outlet (the last control volume of tube row 1) are shown. It is important to note that the refrigerant enters the heat exchanger at tube row 2, as tube row 1 is facing inlet air flows (see Figure \ref{fig: heat exchanger discretization}). It can be seen that more frost accumulates in the control volume of tube row 1 than tube row 2. This is because the moist air first flows through tube row 1 where moisture content is greater than that of tube row 2 since a portion of water vapor carried in the air deposits on tube row 1. In addition, no frost behavior is observed in the control volume at the heat exchanger outlet due to the fact that the superheated refrigerant brings the coil surface temperature above the local dew point temperature of the moist air that results in dry coil operations there. 

\section{Conclusions}\label{sec: conclusion}
This paper presents a novel film volume model that integrates frost and retained/water ice behaviors using an enthalpy method. The film model captures possible operating modes that involve frost formation and melting, and retained water/ice refreezing and melting under cyclic frosting and defrosting operations. The advantage of using the enthalpy method to model these phase change behaviors is that it features a uniform model structure with switched source terms for various operating modes. A switching algorithm based on the Fuzzy modeling approach is proposed to avoid discontinuities under mode switches. The developed film volume model is incorporated into a discretized outdoor flat-tube heat exchanger model which is then coupled to other component models of an automotive heat pump system. The system model is simulated to investigate system responses under cyclic heating and reverse-cycle defrosting operations from a startup. Simulation results indicate that the developed model is capable of capturing the complicated cyclic transients of the refrigerant, air, frost, and retained water/ice in case of an insufficient defrost cycle. The proposed modeling approach can be extremely useful for improved defrost control designs and evaluations of air-source heat pumps as well as low-temperature refrigeration applications. 

Future research efforts can focus on experimentally validating the model in scenarios of retained water/ice presences and utilizing the model to investigate the overall impact of retained water refreezing and melting on system performance over multiple defrost cycles to aid performance improvements.

\nomenclature[P]{$M$}{Mass $[\mathrm{kg}]$}
\nomenclature[P]{$A$}{Area $[\mathrm{m^2}]$}
\nomenclature[P]{$T$}{Temperature $[\mathrm{K}]$}
\nomenclature[P]{$\dot{Q}$}{Heat transfer rate $[\mathrm{W}]$}
\nomenclature[P]{$\dot{m}$}{Mass transfer rate $[\mathrm{kg/s}]$}
\nomenclature[P]{$\delta$}{Thickness $[\mathrm{m}]$}
\nomenclature[P]{$\rho$}{Density $[\mathrm{kg/m^3}]$}
\nomenclature[P]{$\epsilon$}{Frost porosity $[-]$}
\nomenclature[P]{$H$}{Enthalpy $[\mathrm{J}]$}
\nomenclature[P]{$h$}{Specific enthalpy $[\mathrm{J/kg}]$}
\nomenclature[P]{$V$}{Volume $[\mathrm{m^3}]$}
\nomenclature[P]{$\dot{H}$}{Enthalpy flow rate $[\mathrm{W}]$}
\nomenclature[P]{$u$}{Specific internal energy $[\mathrm{J/kg}]$}
\nomenclature[S]{$wv$}{Water vapor}{}
\nomenclature[S]{$wi$}{Retained water/ice}{}
\nomenclature[S]{$w$}{Liquid water}{}
\nomenclature[S]{$f$}{Frost}{}
\nomenclature[S]{$in$}{Inlet}{}
\nomenclature[S]{$out$}{Outlet}{}
\nomenclature[S]{$r$}{Refrigerant}{}

\printnomenclature

\printbibliography
\end{document}